\documentclass[3p]{elsarticle}

\usepackage{hyperref}

\journal{Control Engineering Practice}

\usepackage{amsmath} 
\usepackage{subfig}   
\usepackage[dvipsnames]{xcolor}  
\usepackage{amssymb}

\newenvironment{centry}{    
 \begin{list}{$\bullet$}{
 \settowidth{\labelwidth}{~}%
 \setlength{\leftmargin}{\labelwidth}%
 \addtolength{\leftmargin}{\labelsep}
 \setlength{\parsep}{0cm}%
 }
}
{\end{list}}

\begin{document}

\begin{frontmatter}

\title{AssistMe: Policy iteration for the longitudinal control of a non-holonomic vehicle}

\author{Catalin Stefan Teodorescu\corref{mycorrespondingauthor}}
\author{Tom Carlson}
\address{Aspire Create, University College London, Royal National Orthopaedic Hospital, HA7 4LP, UK}
\cortext[mycorrespondingauthor]{Corresponding author}

\begin{abstract}
In this article we design a physically-inspired model-based assist-as-needed semi-autonomous control (ASC) algorithm to address the problem of safely driving a vehicle (a power wheelchair) in an environment with static obstacles. Once implemented online, the proposed algorithm requires limited computing power and relies on pre-computed (offline) maps (look-up tables). These are readily available by implementing policy iteration that minimizes the expected time to termination (safely stopping near an obstacle), by taking into account: (i) the vehicle dynamics; (ii) the drivers' intention modeled as three separate stochastic processes. We call them the expert driver, the naughty child and the blind driver models.

A study with healthy participants confirmed that ASC outperforms a baseline rule-based control (a statistically significant result).
\end{abstract}


\end{frontmatter}

\section{INTRODUCTION}
This work is inspired by a real need for bespoke control solutions applicable to medical robotics. In the European project ``ADAPT'', clinicians (medical doctors) identified the needs of wheelchair users \cite{LeFrDe2020} and in response, roboticists proposed to make use of technology and novel scientific approaches to tackle those challenges, one of which is addressed in this article.

\subsection{Motivation}
In particular, in health and social care, about 840,000 people in the UK are regular wheelchair users \cite{NHS21a} living with multiple impairments (physical, cognitive, visual, hearing) who might benefit from an assisted robotic wheelchair that can help them navigate safely in any indoor or outdoor environment. There are currently no such commercially available devices on the market. Consequently, they are at risk of bumping unintentionally into obstacles, hurting themselves or the people around them. Moreover, over the last half century, the UK has seen an increase in life expectancy and this trend is expected to continue \cite{NHS21b}. To meet the needs of an ageing society, research in innovative assistive robotic technologies has the potential to empower healthy aged people towards living an independent and active life, while keeping them safe and socially connected \cite{DeAgBi2020,VoLeHa2021}; thus increasing their self-esteem. The beneficiaries of such technology account for 58\% of the people aged 60 and over in UK who suffer from mobility difficulties \cite{AgeUK21}. In 2014 there were 14.9 million people aged 60+ and their number is expected to grow to 18.5 million by 2025 \cite{NHS21c}.
%
%
In a new world reshaped by the current pandemic, novel control solutions may: (i) make life easier for these categories of people who need flexibility; (ii) help health organizations like the National Health Service (NHS) in the UK to overcome issues like limited staff resources.

\subsection{Shared control}
Assist-as-needed \cite{ShBeSa2020} or shared control \cite{AbCaMu18} is a concept involving collaboration between a human and a machine. The human operator informs the machine about their objective (e.g. by actuating a joystick) and the machine interprets that action, then implements it in an optimal way. The algorithms studied in the literature can be divided in two categories, depending on the need to model system dynamics. 
On the one hand there are the \textit{black-box} control solutions like the Probabilistic Shared Control \cite{EzTrDe2017}, optimal control in \cite{DeViPa2016,NaSpBa2016}, and Dynamic Window Approach \cite{FoBuTh1997}, task-oriented control \cite{VoLeHa2021}. 
Arguably the most interesting feature of these algorithms resides in their generic value: they are most suitable in situations where models of the wheelchair dynamics or the user intention do not exist, and can easily be switched (adapted) from one vehicle type to another one. 
The second category of algorithms rely on \textit{white-box} control solutions that require model identification; they are analyzed next.

\subsection{Model-based control}
Moving vehicles obey physical laws which are well-known and understood (e.g. inertia, gravity, etc.). Consequently, it is natural to create a model integrating the capabilities of the machine. Although models in general are a mere reflection of reality, control-oriented models permit taking into account the uncertainties from the early phase of designing the control.
The challenge is to be able to derive control policies based on the control-oriented model and the limited unpredictable information coming from: (i) the sensors observing the environment (e.g. low-cost time-of-flight sensors that can measure only the 1D distance to obstacles whereas their 3D shape remains unknown), and (ii) the human user (e.g. the joystick interface is a projection of what the user thinks, and their intention). In what follows, we shall make the distinction between the availability of significant amount of sensor data (e.g. measurements at high rate) and the limited information provided by each sensor (e.g. 1D distance to an obstacle is not sufficient to know the precise location of an obstacle in 2D Cartesian space; 2D joystick data is only informative about the immediate intention of the driver, not the long-term intention like getting closer to a table in order to serve their lunch). To compensate for this limited information available via online measurements, recent state-of-the-art research uses stochastic models.

\subsection{A user-centered design}
Taking into account the intention of the user is essential. An assistive machine should complement their intention in such a way that the user feels empowered and can make use of their existing functionality (e.g. physical, cognitive, visual, hearing, etc.) \cite{Nis2002}. Contrary to able-bodied people who might be more in favour of autonomy (think of Tesla cars driving autonomously), the potential users of semi-autonomous assist-as-needed technologies want to maximise their level of control, and thus develop new skills. They favour technology that empowers the user rather than further disabling them by taking away control authority and treating them like a precious piece of cargo. Therefore, instead of relying on generic control algorithms intended to satisfy all users, we shall put forward a method that can be tailored to a specific (individual) user. Each wheelchair user has their own style of driving and any particular assist-as-needed algorithm that works for one user might not necessarily satisfy the needs of a second user. For example, an expert wheelchair user might prefer to cut corners at maximum speed, whereas another driver might prefer a different, more cautious strategy. If we ask them to repeat the same operation over and over, their driving style can be characterized using probabilities.

This observation motivates the use of multiple stochastic user models (see later on section~\ref{sec:stoch-driver-models}). 

\subsection{Stochastic Dynamic Programming (SDP)}
In view of making bespoke informed decisions, the SDP framework is generous enough as to allow taking into account stochastic uncertainties (namely driver models) from the early phase of designing the control.
SDP is a control design technique that can handle stochastic models by solving a global optimization problem under constraints. The solution is not analytical, instead it is computed numerically, often on high-performance computers able to handle the burden of intensive computations and fairly large matrices, and comes in the form of optimal control policies \cite{Ber2005}. These operations are carried out offline and the outcome, namely a lookup table containing the optimal actions, is implemented online on hardware with limited computational resources (e.g. industrial controllers or single-board computers). For this reason, SDP is  particular appealing to the automotive sector \cite{TeVaDe2017}.

One significant limitation of SDP is the well-known \textit{curse of dimensionality} (exponential increase of the required computation with each new state, control variable or uncertainty) \cite{Ber2005}. To summarize, SDP does not scale well to problems involving many variables where exact solutions become computationally intractable \cite{HeWaMe2020}. To mitigate its effect: (i) compact-sized problems (e.g. system dynamics with no more than two state variables) are preferred over larger ones; (ii) instead of relying on the \textit{naive} implementation of Bellman's principle, faster algorithms like the \textit{policy iteration} can be used.

\subsection{The control problem}
\label{sec:ctrl-pb}
\begin{figure}[htbp]
   \centering
   \includegraphics[scale=.75, trim=0ex 0ex 32ex 0ex, clip]{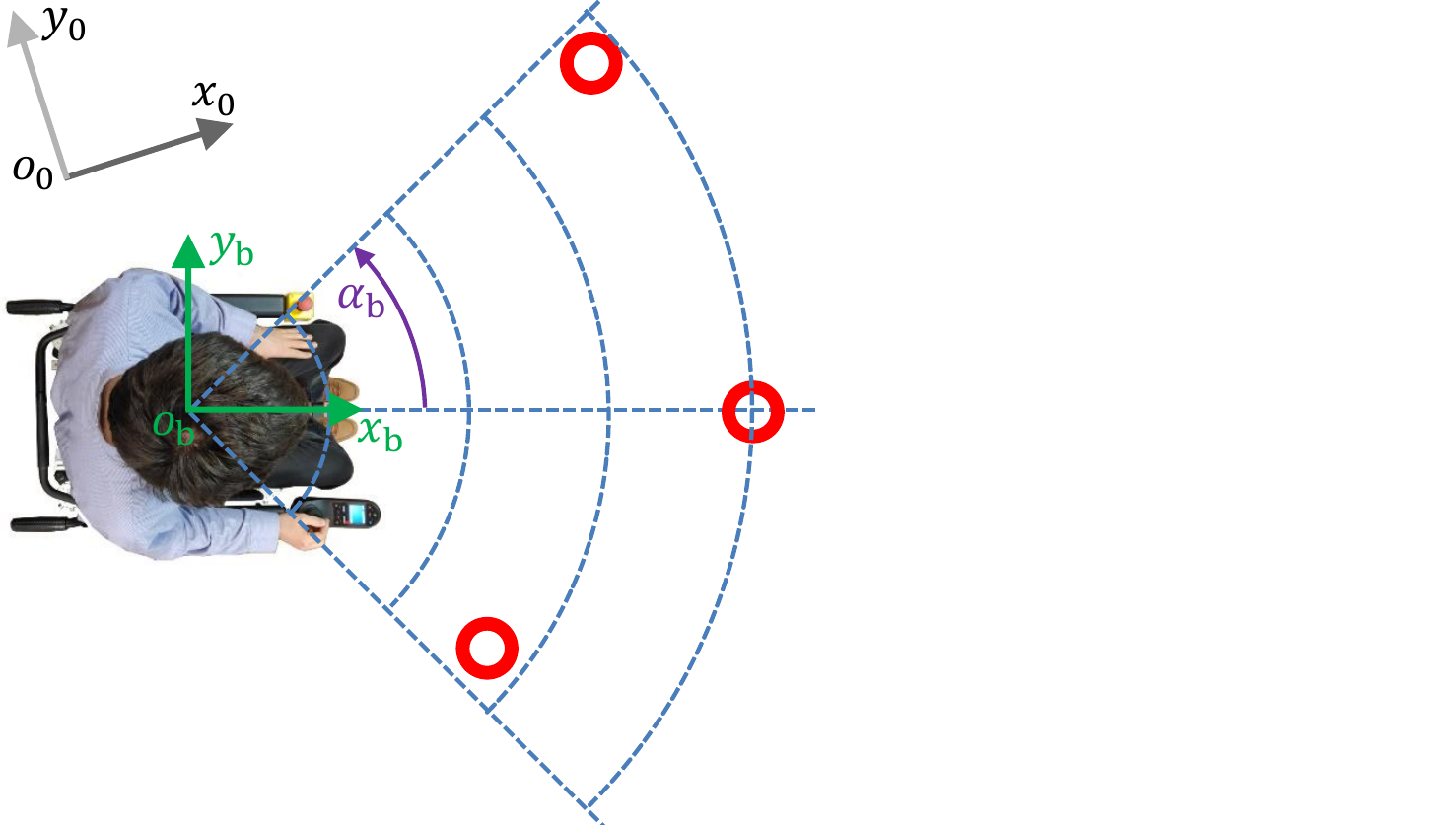}
   \caption{View from above: instrumented wheelchair advancing in an environment with obstacles (in \textcolor{red}{red}) }
   \label{fig:whc_multi_obstac}
\end{figure}

In this article we address the safe-driving problem of a \textit{semi-autonomous} vehicle (a power wheelchair) in an environment with static obstacles, as depicted in Fig~\ref{fig:whc_multi_obstac}. This is still an active area of research \cite{NaSpBa2016}, unlike the purely \textit{autonomous} vehicles case, which is out-of-scope for the purpose of this article and where robust and well-established path planning algorithms are readily available \cite{LiLiTa2020}, \cite[\S 11.1.5]{PyChJu2017}, \cite{BaBaCe2018}. 

\subsection{Our contribution}
In this article we propose a novel shared control algorithm based on policy iteration.
To the best of our knowledge, policy iteration has not been proposed yet in the scientific literature for addressing the problem stated in section~\ref{sec:ctrl-pb}. It is our intention to make a significant contribution in this setting as follows. 

Inspired from the automotive sector, where 
a rather limited number of successful implementations of policy iteration algorithms have been published (see, e.g. \cite{JoAsBo2007,LiXuHa2021}), this article builds upon the seminal work \cite{LiPeGr2004}. Our article is similar in the sense that we shall reuse the same idea of implementing policy iteration to compute global optimal policies in conjunction with stochastic models.
However, our contribution differs with respect to the following points: (i) the application is different (a power wheelchair, not a passenger car); (ii) the optimization is carried out in terms of total time to complete the task (obstacle avoidance), not energy (e.g. minimize fuel consumption \textit{versus} battery usage in hybrid electric vehicles  \cite{LiPeGr2004,LiWaHa2018,LiXuHa2021,TeVaDe2017}); (iii) the framework is different (control of semi-autonomous vehicles, not autonomous vehicles); (iv) the stochastic model is different (user intention, not the driving cycle); (v) environment awareness is ensured online via time-of-flight sensors able to detect obstacles, whereas other means are necessary to identify (classify) driving cycles.

The model-based control solution we propose can be tailored to reach a high level of customization (personalisation) in terms of a specific driving style of any individual user. 

\subsection{Data-driven control}
Among other candidates able to solve the problem stated in section~\ref{sec:ctrl-pb}, \textit{reinforcement learning} was excluded for the following reasons. Although it has the advantages of using a theoretical framework that is more general in contract to optimal control theory \cite{DePeRa2008}, and the ability to learn complex nonlinear behaviors \cite{SuBa2018}, there are other properties which seem detrimental to our problem. In particular, it relies on the \textit{exploitation versus exploration} paradigm which is not suited for safe-critical applications; it requires a prohibitively large number of training iterations before getting to an effective (operational) result \cite{OsScBa2016}.

\section{Policy Iteration}
In this section we formulate an infinite horizon time-optimal stochastic shortest path problem \cite[\S 7.2]{Ber2005}. This is a finite-state problem involving a finite (but rather large) state space. 
The framework was largely inspired by the the inventory control problem in \cite[Ex. 1.3.2, p. 28]{Ber2005}.
We start by presenting the modeling of the vehicle dynamics followed by the driver models, then set forth the control design. The aim is to compute a map (or lookup table) that is informative about the expected time to hit an obstacle for each particular stochastic driver model. We will use this information in the next section~\ref{sec:ASC} to elaborate an assist-as-needed semi-autonomous control algorithm.

\begin{figure}[htbp]
\centering
\vspace{1ex}
\subfloat[view from above]{
 \mbox{  \includegraphics[scale=.62, trim=36ex 0ex 0ex 0ex, clip]{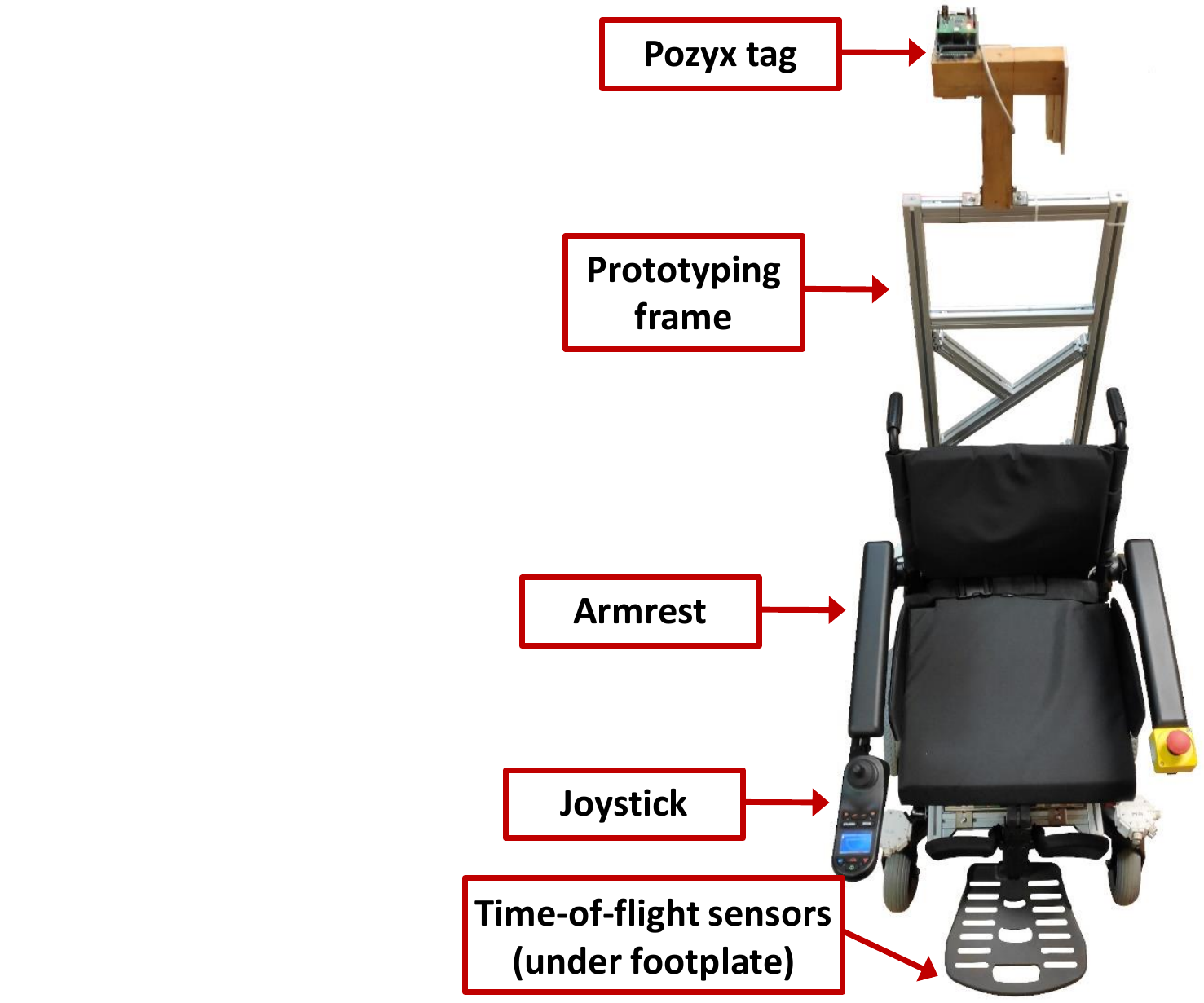} } 
 \label{fig:whc_pos_track}
} %
\hfill
\subfloat[view under the seat where bespoke electronics was added]{ 
 \mbox{ \includegraphics[scale=.62, trim=0ex 0ex 8ex 0ex, clip]{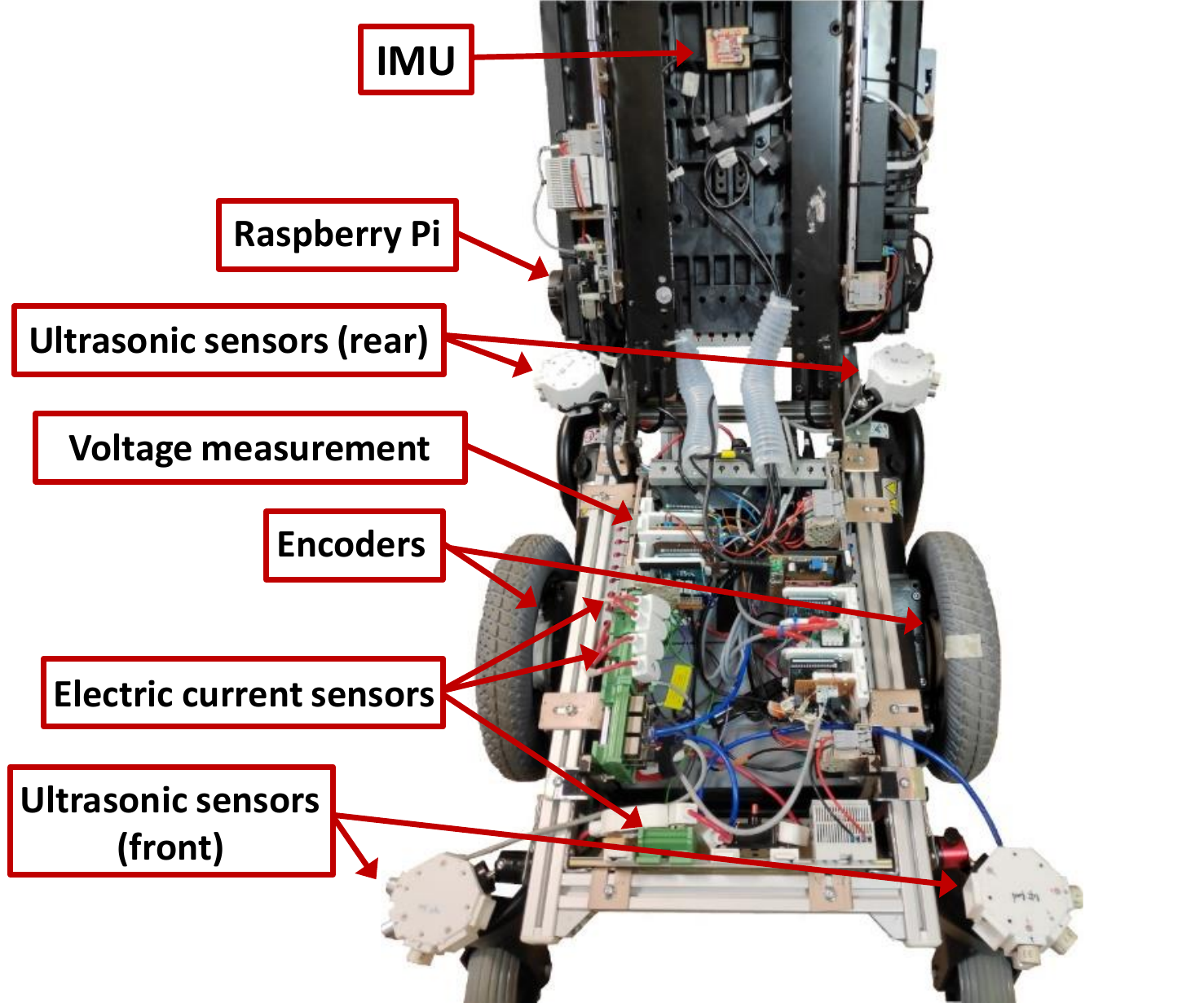} } 
 \label{fig:whc_under_seat} 
} %
\caption{An off-the-shelf commercial wheelchair converted into a prototype instrumented semi-autonomous robotic vehicle}
\label{fig:instrum-whc}
\end{figure}

\subsection{Vehicle model}
Wheelchair models are known in the literature \cite{FeGuBu2021}. Here, we focus on the dynamics of the instrumented wheelchair from Fig.~\ref{fig:instrum-whc}, which are governed by the following system equations \cite{TeZhCa2020a}:
\begin{subequations}
\begin{align}
x_{k+1} &= \max (0, \min(d_\text{max}, \,x_k + v_k \Delta{}t )) \label{eq:dyna-model-straight-line-eq1} \\
v_{k+1} &= \max (v_\text{min}, \,\min (v_\text{max}, \sigma_1 v_k + \sigma_2 (v_{\text{d},k}+v_{\text{u},k} )) \label{eq:dyna-model-straight-line-eq2}
\end{align} 
\label{eq:dyna-model-straight-line}
\end{subequations}
with the constraint that $v_{\text{u},k}$ is such that 
\begin{align}
v_\text{min} <= v_{\text{d},k}+v_{\text{u},k} <= v_\text{max} 
\label{eq:vreq-constraint}
\end{align}
All variables and parameters are defined in Table~\ref{tab:nomencl} together with their units of measure; subscript $k$ indicates the time index. The physical meaning of the constraint \eqref{eq:vreq-constraint} is to avoid requesting an input velocity $v_{\text{d},k} + v_{\text{u},k}$ that is more than the wheelchair's power module profile is set to handle. 
The physical meaning of the saturation in:
\begin{centry}
\item eq.~\eqref{eq:dyna-model-straight-line-eq1} is that we are concerned with vehicle motion towards the nearest obstacle located in the field-of-view of a sensor (e.g. an ultrasonic or time-of-flight sensor).
\item eq.~\eqref{eq:dyna-model-straight-line-eq2} is that each motor servo-drive (power module) includes a low-level safety functionality which bypasses (overrides) any requested control action beyond $v_\text{max}>0$ and below $v_\text{min}<0$.  
\end{centry}

\begin{table}[htbp]
\caption{Nomenclature}
\label{tab:nomencl}
\begin{center}
\begin{tabular}{|p{.8cm}|p{5.6cm}|p{.7cm}|}
\hline
Symbol & \centering Meaning & Units  \\ \hline
$x_k$ & position of the vehicle expressed in coordinates of the fixed frame 0 i.e. $x_k\triangleq{}x_{o_\text{b}, k}^0$ at time instant $k$ & m \\ \hline
$d_\text{max}$ & maximum distance an obstacle can be detected (perceived) by a sensor,  expressed in coordinates of the moving base frame i.e. $d_\text{max}\triangleq{}d_\text{o}^\text{b,max}$ & m \\ \hline
$v_k$ & linear velocity of the vehicle at time instant $k$ & m/sec \\ \hline
$v_\text{min}$, $v_\text{max}$ & minimum and maximum linear velocity, respectively & m/sec \\ \hline
$v_{\text{d},k}$ & driver's demand (or intention) at time instant $k$ & m/sec \\ \hline
$v_{\text{u},k}$ & supervisory control variable at time instant $k$ & m/sec \\ \hline
\hline
$\Delta{}t$ & sampling time & sec \\ \hline
$\sigma_1$, $\sigma_2$ & physically-inspired experimentally identified parameters \cite{TeCaEb2021} & -- \\ \hline
$\alpha_\text{b}$ & field of view (sector area) of sensors detecting obstacles in Fig.~\ref{fig:whc_multi_obstac} & rad \\ \hline
\end{tabular}
\end{center}
\end{table}

\subsection*{Notations}
Consistent with notations in \cite{SpHuVi2020}, superscripts ``0'' and ``b'' in Table~\ref{tab:nomencl} indicate that that quantity was expressed in coordinates of the inertial frame and base frame, respectively. Subscripts indicate the object they refer to, e.g. ``o'' stands for obstacle.
In Fig.~\ref{fig:whc_multi_obstac}, $o_0x_0y_0$ represents the fixed (inertial) frame and $o_\text{b}x_\text{b}y_\text{b}$ is the moving (base) frame. The latter has its origin at midpoint between the two main wheels, and the $x_\text{b}$-axis points in the direction of the vehicle advancing forward.

\subsection{Stochastic driver models}
\label{sec:stoch-driver-models}
\begin{figure}[htbp]
   \centering
   \includegraphics[scale=.75, trim=8ex 8ex 12.2ex 24ex, clip]{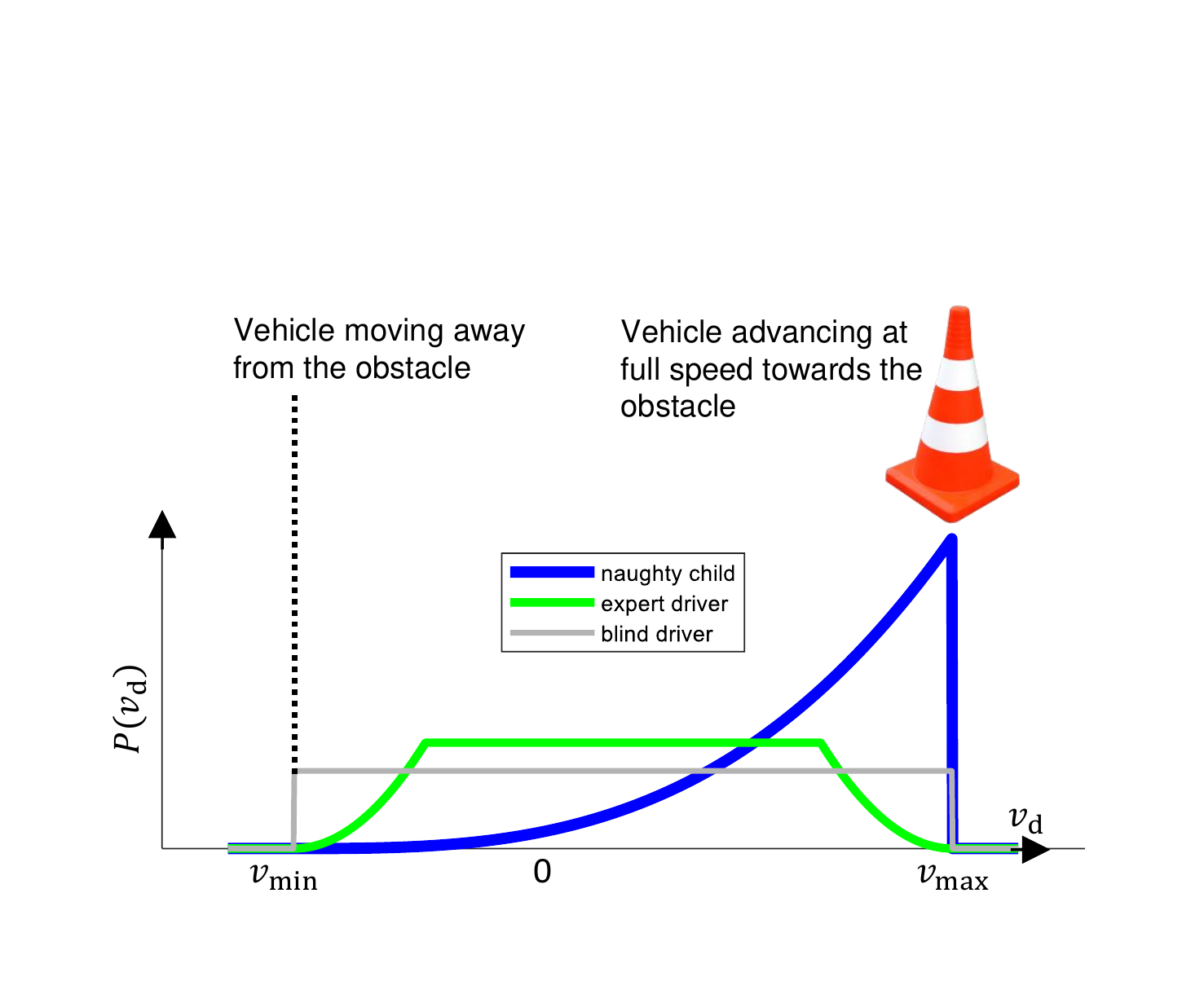}
   \caption{Conceptual illustration of the probability distribution of three stochastic driver models}
   \label{fig:whc_stoch_driver_models}
\end{figure}

In this work, we define three \textit{generic} stochastic user models, which we call the \textit{expert driver}, the \textit{blind driver} and the \textit{naughty child}, depicted in Fig.~\ref{fig:whc_stoch_driver_models}. 
The \textit{naughty child} has tendency to accelerate towards the obstacle (with the intention to hit it as means to explore the world), whereas the \textit{expert driver} carefully maneuvers the vehicle as to avoid getting in the situation of hitting the obstacle; finally, the \textit{blind driver} is unable to acknowledge the presence of the obstacle and makes no difference in terms of driving style whether the obstacle is present or not (thus we used an uniform distribution).  

We end this section by mentioning that fine tuning or learning these driver models is out-of-scope and will be subject to future research. Inspired by \cite{MaGoHa2020}, we intend to use \textit{Gaussian processes} - a powerful tool in \textit{machine learning}.

\subsection{The control problem}
\begin{figure}[htbp]
   \centering
   \includegraphics[scale=.5, trim=0ex 0ex 0ex 0ex, clip]{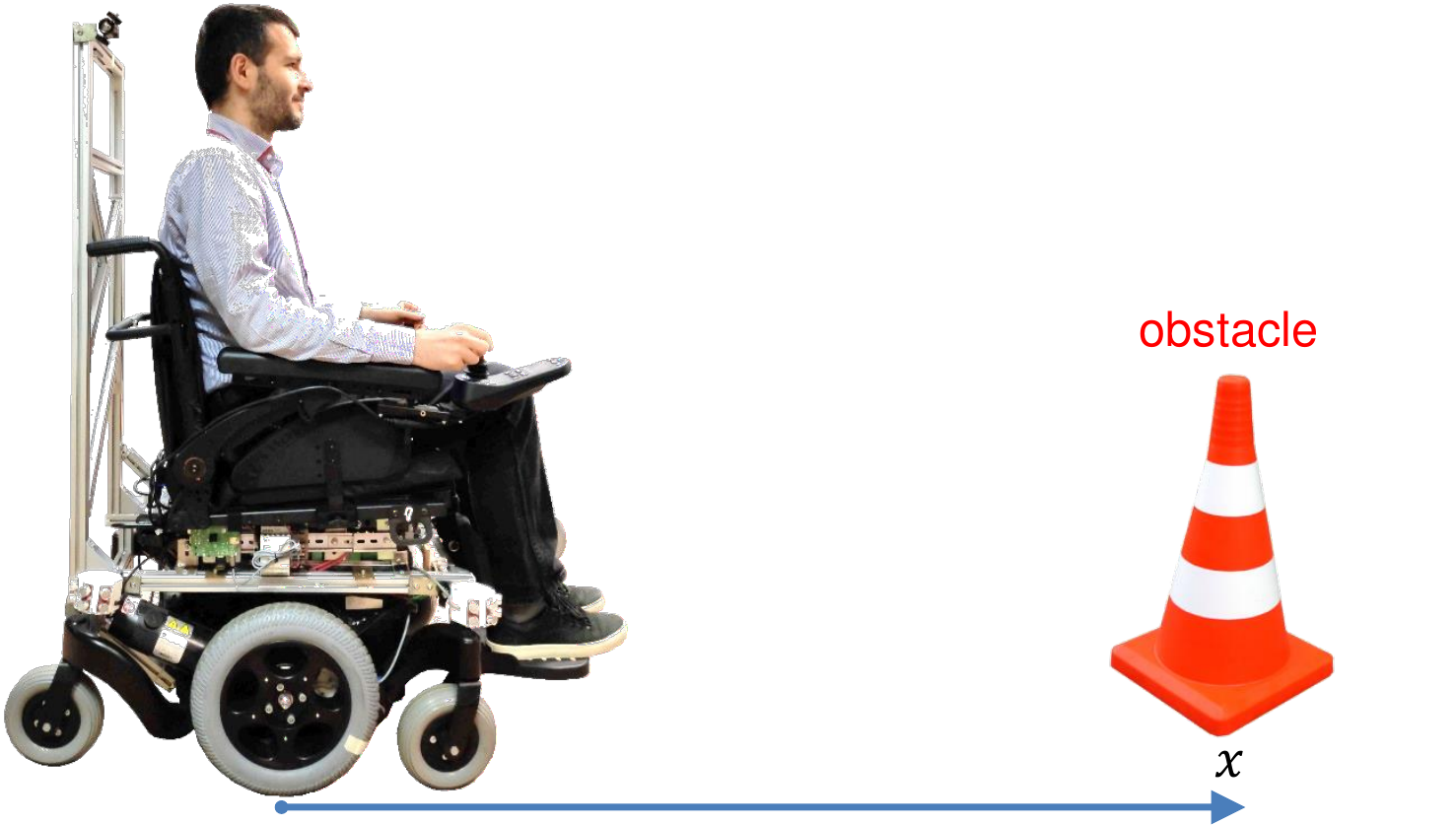}
   \caption{Vehicle advancing in straight line with an obstacle located in front}
   \label{fig:whc_side-view}
\end{figure}

In this section, we start by analyzing the simpler case depicted in Fig~\ref{fig:whc_side-view} of a vehicle advancing in straight line with an obstacle located in front of it ($\alpha_\text{b}=0$ in Fig.~\ref{fig:whc_multi_obstac}). Later on, in section~\ref{sec:ASC-2D} we present an extension towards the more general case of a vehicle able to detect multiple obstacles on a sector area in front of it ($\alpha_\text{b}>0$ in Fig.~\ref{fig:whc_multi_obstac}).

For each stochastic driver model, we want to know the minimum expected \textit{time to termination}, defined as the time elapsed from the initial state of the moving vehicle $(x,v)$, where notations $x\triangleq{}x_{k=0}$ and $v\triangleq{}v_{k=0}$ correspond to the initial (or starting) instant, to the moment where it stops safely near that obstacle, without hitting it. In other words, we address the following optimization problem:  
\begin{align}
J(x,v) = \min_{v_{\text{u},k}} \,\mathop{\mathbb{E}}_{v_{\text{d},k}} \left\lbrace \sum_k g(x_k,v_k) \right\rbrace
\label{eq:def-J}
\end{align}
along trajectories defined by \eqref{eq:dyna-model-straight-line} and such that constraint \eqref{eq:vreq-constraint} holds; $k=1,..,\infty$ is the number of stages. In \eqref{eq:def-J}, $\mathbb{E}$ is the \textit{expected value} operator expressing a sum of all possible (feasible) trajectories that are all weighted together by their probability of occurrence associated to uncertainties $v_{\text{d},k}$. 
The policy iteration algorithm that implements the cost \eqref{eq:def-J} will compute that control policy which works best for all feasible trajectories (in the sense of a weighted sum). However, the dominant trajectories that have higher weights will account the most in the process of computing the optimal control policy.
In \eqref{eq:def-J}, $g(\cdot,\cdot)$ is the cost per stage, to be defined next. 

\subsubsection{Cost per stage without penalty}
\begin{figure}[tbp]
\centering
\vspace{1ex}
\subfloat[without additional penalty]{
 \mbox{  \includegraphics[scale=.75, trim=31ex 54ex 35ex 55ex, clip]{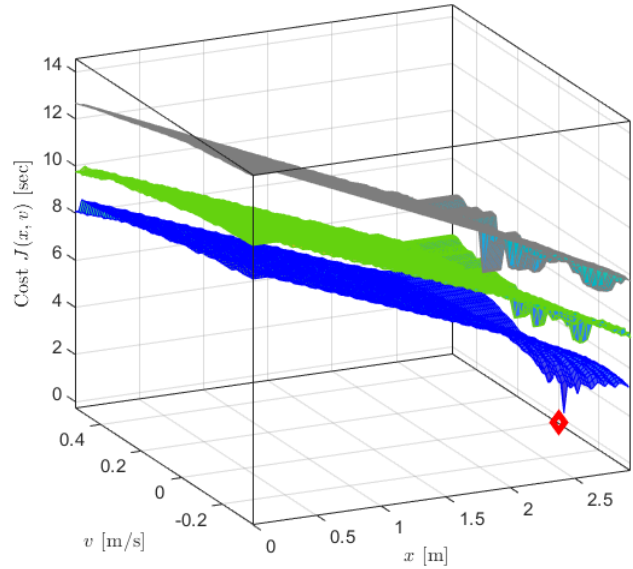} } 
 \label{fig:whc_policy-iter-algo-cost-fcts-woPenalty-all}
} %
\hfill
\subfloat[with additional penalty]{ 
 \mbox{ \includegraphics[scale=.75, trim=29ex 53ex 34.5ex 58ex, clip]{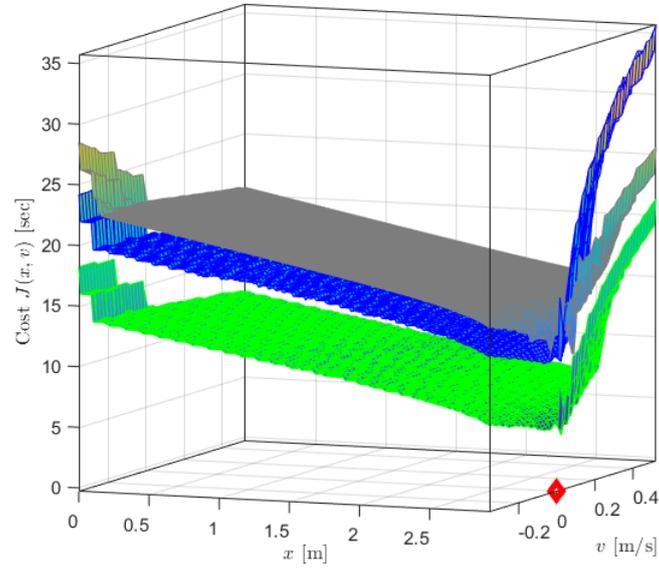} } 
 \label{fig:whc_policy-iter-algo-cost-fcts-wPenalty-all} 
} %
\caption{Results of running the policy iteration algorithm with or without additional penalty on the cost per stage $g(\cdot)$: the total cost represents the minimum expected time to reach the terminal state (in \textcolor{red}{red}), for three stochastic driver models: expert driver (in \textcolor{green}{green}), naughty child (in \textcolor{blue}{blue}), blind driver (in \textcolor{gray}{gray})}
\label{fig:whc_policy-iter-algo-cost-fcts}
\end{figure}

By choosing $g(\cdot,\cdot) \equiv \Delta{}t$ in \eqref{eq:def-J}, as expected with solving time-optimal problems, the control policy comes in the form of a bang-bang control \cite{BoRo2005}. 
It represents an aggressive maneuver of that specific stochastic driver model advancing towards the obstacle, but still being able to stop safely near the obstacle, without bumping into it. An interesting fact of this formulation is that the total cost \eqref{eq:def-J} has a physical meaning: the map in Fig.~\ref{fig:whc_policy-iter-algo-cost-fcts-woPenalty-all} shows the time measured in seconds for executing this maneuver by the 3 stochastic driver models. 

The interpretation of results in Fig.~\ref{fig:whc_policy-iter-algo-cost-fcts-woPenalty-all} follows. Note the descending slope as the vehicle approaches the terminal state. Comparing the 3 drivers' results, we see that, as expected, it takes more time in average for the \textit{blind driver} to reach the terminal state compared to the other driver models. The \textit{naughty child} outperforms the other two drivers in reaching the terminal state due to the tendency of intentionally wanting to hit the obstacle. The \textit{expert driver}'s results are located in-between the other two users.  

\begin{figure}[htbp]
   \centering
   \includegraphics[scale=.68, trim=27ex 66ex 30ex 70ex, clip]{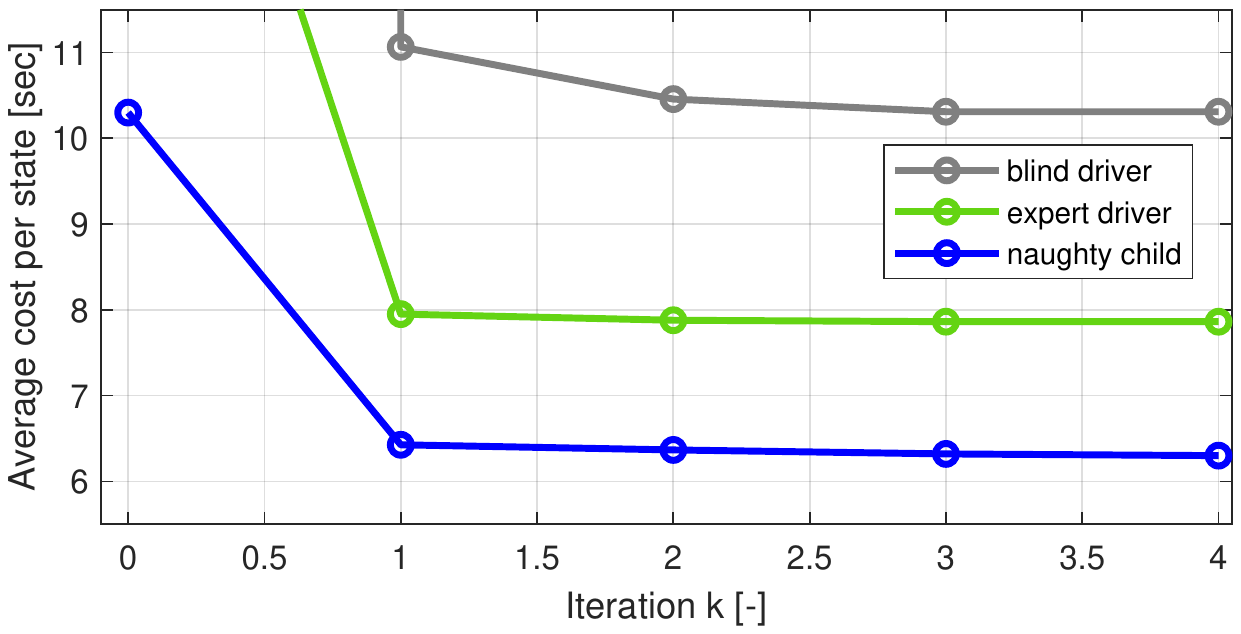}
   \caption{Convergence of the policy iteration algorithm (the first 5 iterations)}
   \label{fig:whc_policy-iter-conver}
\end{figure}

In Fig.~\ref{fig:whc_policy-iter-conver} we study the convergence of the policy iteration algorithm. At each iteration $k$, we computed the average of the map of the total cost (examples of such maps are illustrated in Fig.~\ref{fig:whc_policy-iter-algo-cost-fcts-woPenalty-all}). It can be seen that after $k=4$ iterations, the slope of the convergence curve is almost zero and the policy iteration algorithm can be stopped.

The policy iteration algorithm requires a finite number of states. The saturations defined in \eqref{eq:dyna-model-straight-line} motivate the usage of bounds for the $(x,v)$-grid (see the axes of Fig.~\ref{fig:whc_policy-iter-algo-cost-fcts-woPenalty-all}); the constraint \eqref{eq:vreq-constraint} suggests bounds for the $(v_\text{d},v_\text{u})$-grid. By carefully investigating the resulting grid, a problem arises in two situations: (i) $x=0$ and $v<0$; (ii) $x=d_\text{max}$ and $v>0$. In both cases, what happens physically is that the vehicle would normally advance outside of the $x$-grid. To cope with this situation, we defined an environment similar to a toy circuit where a toy vehicle bumps into damper walls physically limiting its ability to advance any further (beyond the boundaries of the circuit) and also protecting the toy from damage. Next, we show how to remove this artificial limitation by introducing soft constraints.

\subsubsection{Cost per stage with penalty}
\begin{figure}[htbp]
   \centering
   \includegraphics[scale=.66, trim=0ex 4ex 40ex 37ex, clip]{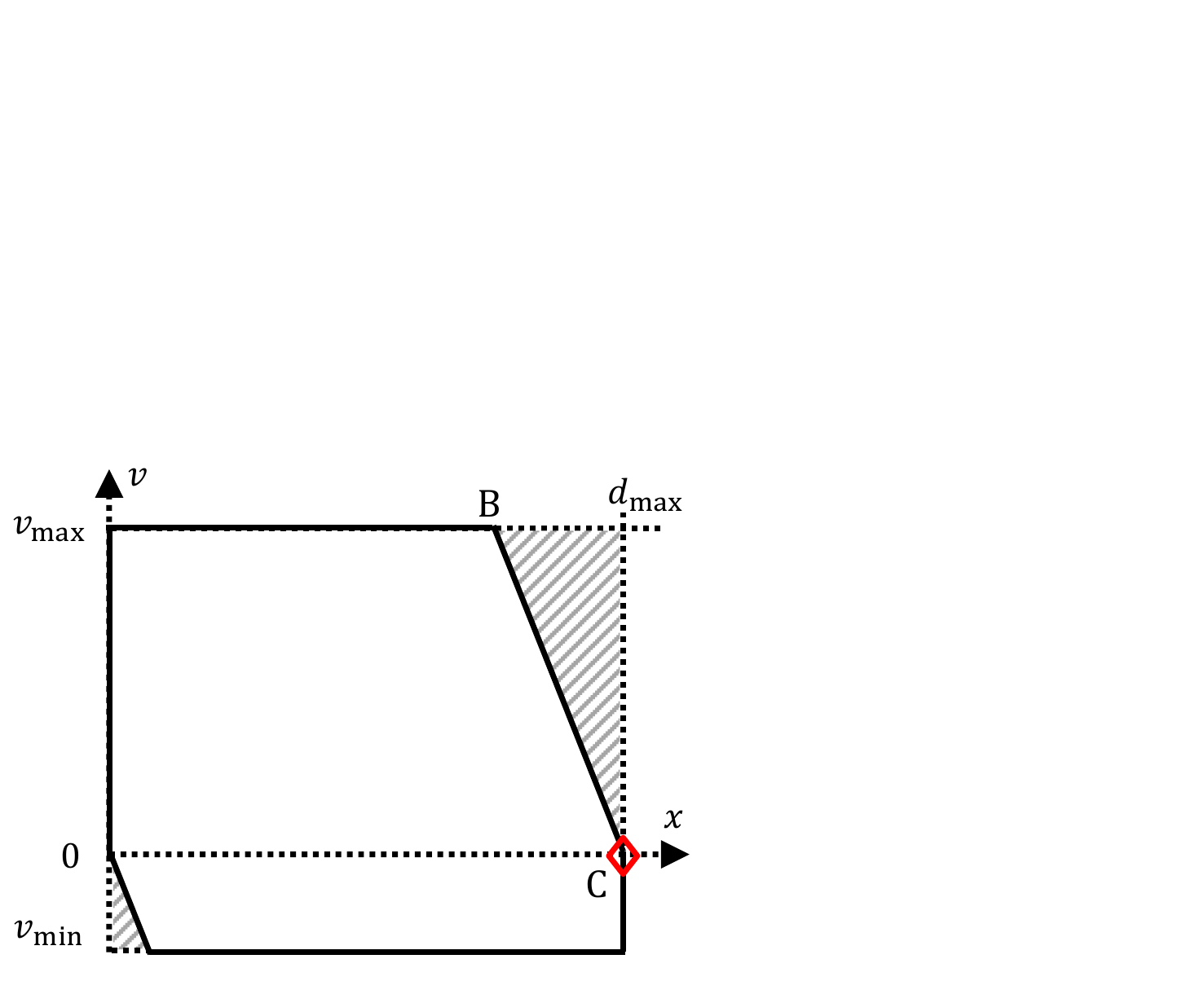}
   \caption{Conceptual drawing of areas posing safety concerns (shaded) \textit{versus} a collision-free area (white)}
   \label{fig:whc_area_safety_pbs}
\end{figure}

By choosing
\begin{align}
g(x_k,v_k) = \left\lbrace \begin{array}{ll} \beta \Delta{}t, & \text{if } (x_k=0 \text{ and } v_k<0) \text{ OR} \\
& \quad (x_k=d_\text{max} \text{ and } v_k>0) \\
 \Delta{}t, & \text{otherwise}
						  \end{array}
             \right.
\label{eq:def-g_wPenalty}
\end{align}
with $\beta\gg0$ a penalty factor, the cost function in \eqref{eq:def-J} changes both its mean value as well as the shape: see Fig.~\ref{fig:whc_policy-iter-algo-cost-fcts-wPenalty-all}. 
First, we shall analyze the shape.
Interestingly, we can now easily identify states where the vehicle would crash into the obstacle (the area around $x=d_\text{max}$ and $v>0$) or would advance backwards beyond the range of sight of the sensors ($x=0$ and $v<0$): they correspond to regions where the cost function surges in Fig.~\ref{fig:whc_policy-iter-algo-cost-fcts-wPenalty-all}. From this we can conclude that the state-space domain can be conceptually sectored as depicted in Fig.~\ref{fig:whc_area_safety_pbs}.
Moreover, it makes sense to think of designing a physically inspired assist-as-needed semi-autonomous control (ASC) within the white sector area. Outside this area, safety measures need to be enforced (e.g. by taking authority from the user, active braking, etc.) which is a topic out-of-scope for this article. In the following, we call the white area in Fig.~\ref{fig:whc_area_safety_pbs} the feasibility region where ASC may be enabled. 

Second, we shall analyze the z-axis values of the maps in Fig.~\ref{fig:whc_policy-iter-algo-cost-fcts-wPenalty-all} compared to their counterpart in Fig.~\ref{fig:whc_policy-iter-algo-cost-fcts-woPenalty-all}.
Unfortunately, in view of designing a physically inspired ASC, we cannot make use of the values of the maps in Fig.~\ref{fig:whc_policy-iter-algo-cost-fcts-wPenalty-all} within the feasibility region: a drawback of having introduced the penalty in \eqref{eq:def-g_wPenalty} is that the mean values of the maps in Fig.~\ref{fig:whc_policy-iter-algo-cost-fcts-wPenalty-all} went up compared to the results in Fig.~\ref{fig:whc_policy-iter-algo-cost-fcts-woPenalty-all}, to the extent of losing their physical meaning. 
To quantify this in an intuitive way, we shall consider the following simple calculation of a vehicle advancing at $v_\text{max}=0.54$ m/sec driven by a deterministic user ($v_\text{d} \equiv v_\text{max}$), unsupervised ($v_\text{u}\equiv0$): it covers the distance from 0 to $d_\text{max}=2.83$ m in about $d_\text{max}/v_\text{max}\approx 5.24$ sec, whereas values in Fig.~\ref{fig:whc_policy-iter-algo-cost-fcts-wPenalty-all} are at least two times higher than that. 

We end this section by mentioning that the cost maps in Fig.~\ref{fig:whc_policy-iter-algo-cost-fcts} could have been computed using the less computationally-intensive \textit{value iteration} algorithm \cite[\S 7.2]{Ber2005}. Instead, we chose to implement a different algorithm: although we did not exploit one of the main benefits of \textit{policy iteration}, which is having access to the computed optimal control actions, this might change in future implementations as a way to quantify how different these actions are with respect to the generic driver model.

\section{Assist-as-needed semi-autonomous control (ASC)}
\label{sec:ASC}
\begin{figure}[tbp]
\centering
\vspace{1ex}
\subfloat[Setting up an arbitrary threshold (in \textcolor{cyan}{cyan}); zoom on the naughty child's map (in \textcolor{blue}{blue}) from Fig.~\ref{fig:whc_policy-iter-algo-cost-fcts-woPenalty-all}, around the terminal state (in \textcolor{red}{red})]{
 \mbox{  
         \includegraphics[scale=.75, trim=36ex 61ex 37ex 60ex, clip]{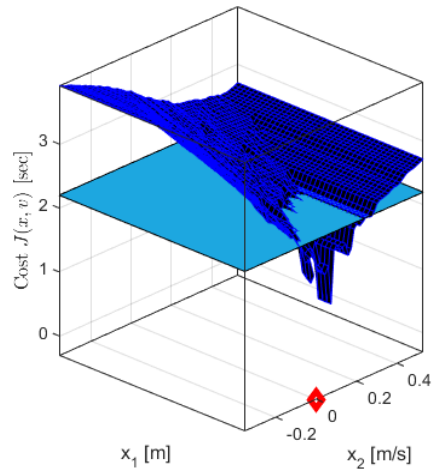} 
      } 
 \label{fig:whc_policy-iter-concept-assist-as-needed-view-lateral}
} %
\hfill
\subfloat[Conceptual drawing in state-space of the working principle of the ASC algorithm: the intention is to reduce any unsafe linear velocity (see e.g. the \textcolor{red}{red} solid circle) within the area of the triangle ABC towards a safe value (see the \textcolor{green}{green} solid circle); note the terminal state is illustrated as a \textcolor{red}{red} rhombus]{ 
 \mbox{ \includegraphics[scale=.66, trim=10ex 4ex 10ex 35ex, clip]{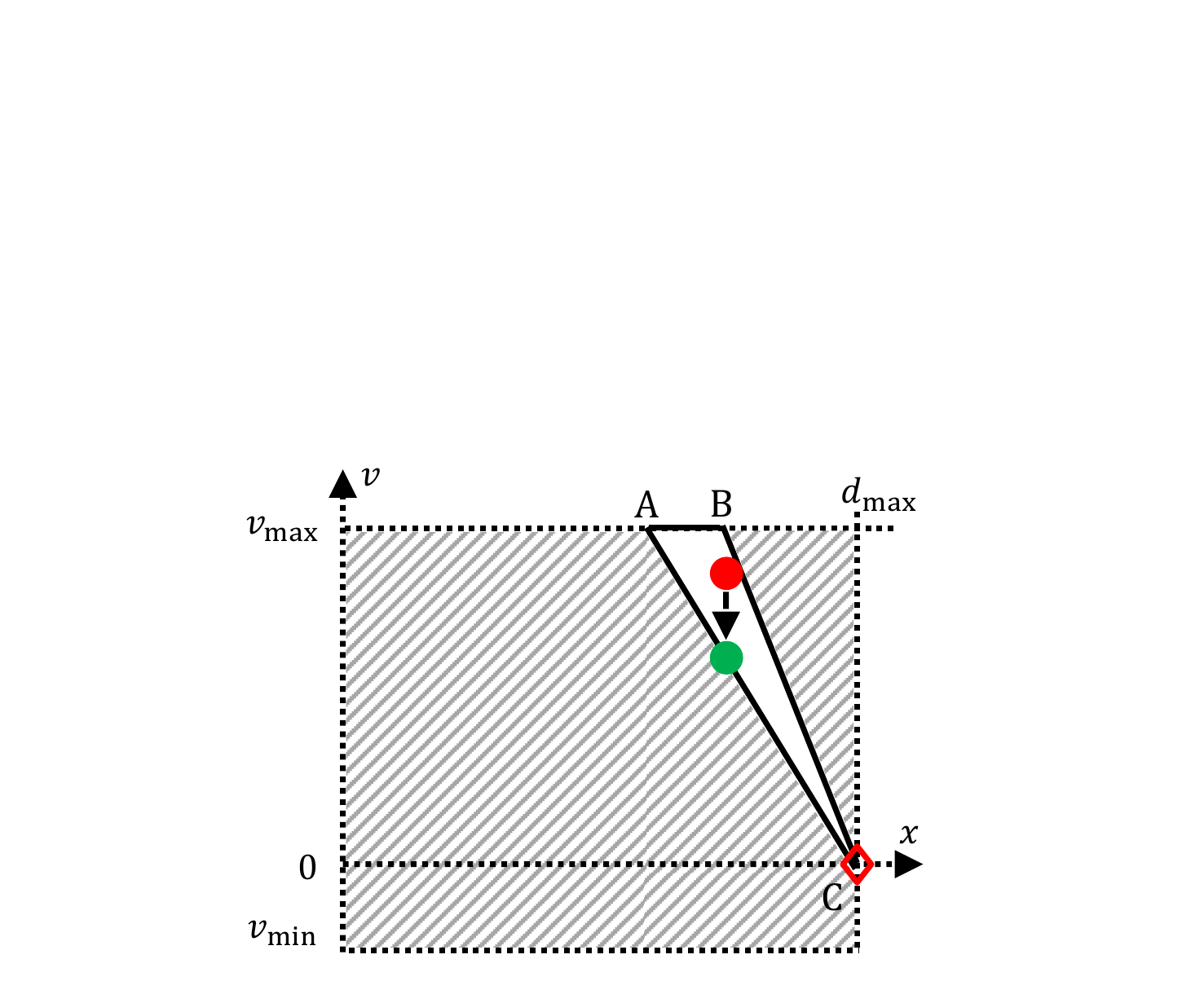} } 
 \label{fig:whc_schematic_ASC} 
} %
\caption{Designing the ASC algorithm}
\label{fig:whc_policy-iter-concept-assist-as-needed}
\end{figure}

In this article we propose to design the ASC as follows. 
We start by setting up an arbitrary threshold representing time to execute a safe stop maneuver for each stochastic driver model (see the plane in Fig.~\ref{fig:whc_policy-iter-concept-assist-as-needed-view-lateral}).
Next, we shall make use of the associated map from Fig.~\ref{fig:whc_policy-iter-algo-cost-fcts-woPenalty-all} within the feasibility region of Fig.~\ref{fig:whc_area_safety_pbs} (the white sector area).
The projection onto the state space of the intersection between this plane and the map, 
gives a new boundary (see the line segment AC in Fig.~\ref{fig:whc_schematic_ASC}) to the region where the intention is to enable ASC (schematically depicted by the triangle ABC in Fig.~\ref{fig:whc_schematic_ASC}).
Motivated by the observation that $v=v_\text{d} + v_\text{u}$ in steady-state in \eqref{eq:dyna-model-straight-line-eq1}, we are now ready to present the guiding principle for driving the vehicle assisted, by modifying the user input data (the joystick).

Say, the \textcolor{red}{red} solid circle in Fig.~\ref{fig:whc_schematic_ASC} illustrates the current condition of the vehicle, expressed in terms of actual velocity $v$ and distance $x$ to an obstacle. Because it is located in the white area where we decide to enable ASC, the driver's actual demand $v_\text{d}$ will be reduced by properly adjusting $v_\text{u}$ to a value corresponding to the \textcolor{green}{green} solid circle. In other words, the \textcolor{green}{green} solid circle corresponds to $v_\text{d} + v_\text{u}$. Reiterating this operation at every time instant (sampling time), the vehicle's actual velocity $v$ will eventually tend towards the boundary of the white area (the line segment AC in Fig.~\ref{fig:whc_schematic_ASC}). Gradually, the vehicle converges safely towards the terminal state.  

Before moving to the next section we would like to mention that generating each map in Fig.~\ref{fig:whc_policy-iter-algo-cost-fcts} required significant computation power, both in terms of running time (about 24 hours of continuous operation on an i7-8750H CPU workstation laptop) and memory (we used all the 32GB RAM available). Possible ways to decrease the running time are suggested in the literature, e.g. in \cite{DePeRa2008} by training a \textit{Gaussian process} (GP) to learn the continuous total cost function by using a coarser state-space grid compared to what we used to compute Fig.~\ref{fig:whc_policy-iter-algo-cost-fcts}; another GP-based approximate DP is presented in \cite{BeMaFu2020}; we leave the implementation of these ideas as future work. 

Next, we shall generalize our results to the more general case depicted in Fig.~\ref{fig:whc_multi_obstac} of a vehicle advancing not only in a straight line, but in 2D Cartesian space, with multiple objects in front of it.

\subsection*{Extension of ASC to the 2D Cartesian case}
\label{sec:ASC-2D}
\begin{figure}[htbp]
   \centering
   \includegraphics[scale=.75, trim=0ex 0ex 32ex 0ex, clip]{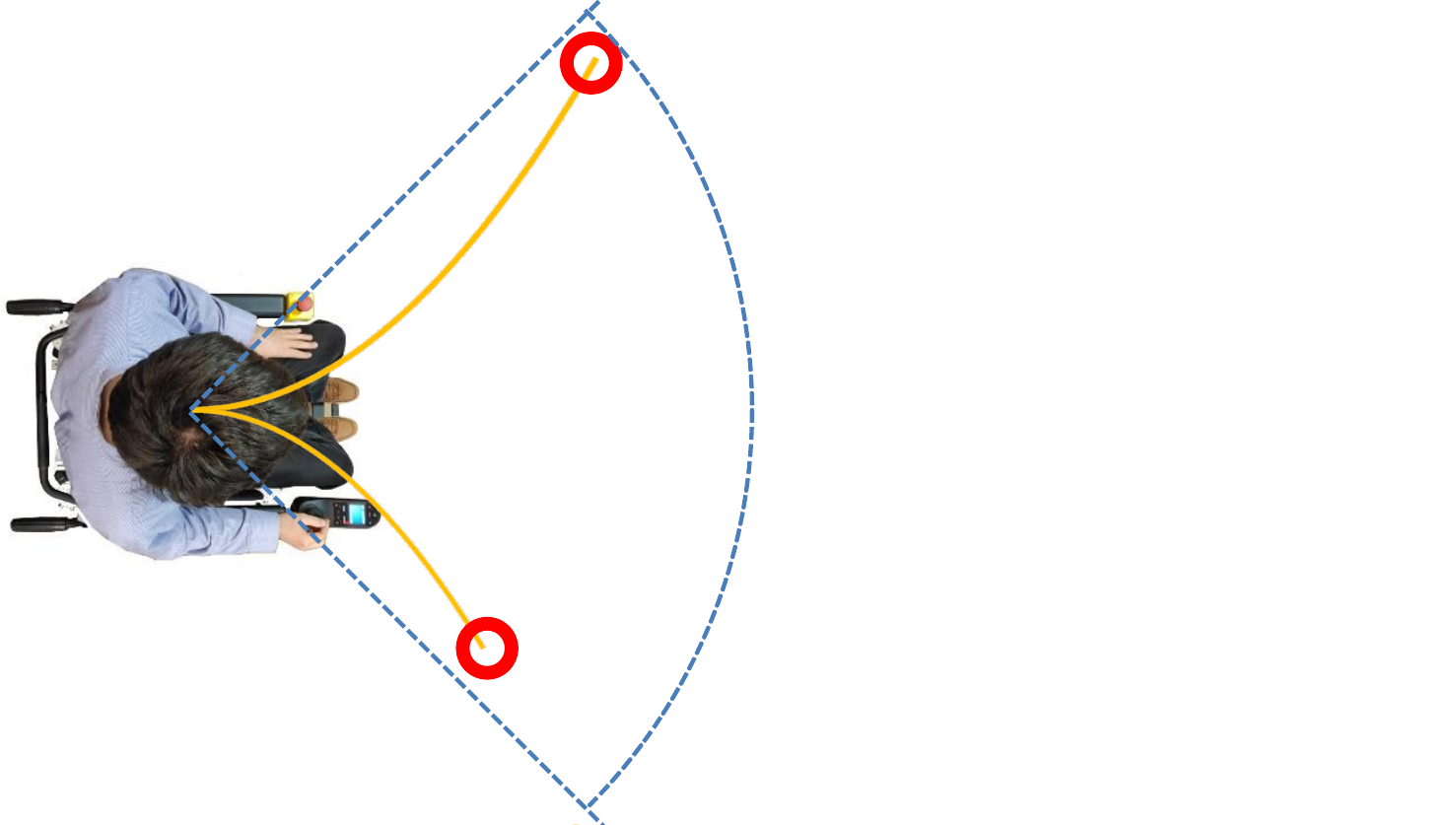}
   \caption{Second order polynomials (in \textcolor{BurntOrange}{amber}) used to approximate the time-optimal trajectory to obstacles (in \textcolor{red}{red})}
   \label{fig:whc_traj_gen_multi_obstacles}
\end{figure}

The same principles put forward in the previous section scale up to the general case of a vehicle able to sense the location of obstacles within a sector area $\alpha_\text{b}>0$ in Fig.~\ref{fig:whc_multi_obstac} as follows. A first attempt would be to extend the system dynamics \eqref{eq:dyna-model-straight-line}--\eqref{eq:vreq-constraint} with additional state variables, then run offline the policy iteration algorithm to generate maps that should provide information about the minimum expected time to advance safely towards each obstacle. 
The problem with this approach is the \textit{curse of dimensionality} mentioned in the Introduction, making it intractable both in terms of computation power (probably in the order of months of continuous CPU operation) as well as memory RAM (in the order of terabytes). 
For this reason we propose to use a second approach where the idea is to reuse the results of the previous section in conjunction with the following assumptions (approximations): 
\begin{centry}
\item the time-optimal trajectory of a vehicle advancing towards an obstacle can be approximated using a polynomial (see Fig.~\ref{fig:whc_traj_gen_multi_obstacles});
\item in 2D Cartesian space, the vehicle dynamics on the time-optimal trajectory towards an obstacle can be approximated using the straight-line motion in 1D Cartesian space; 
in other words there are weak coupling effects between the longitudinal and the lateral motion dynamics of the vehicle. 
\end{centry}

\begin{figure}[htbp]
   \centering
   \includegraphics[scale=.75, trim=0ex 2.5ex 32ex 10ex, clip]{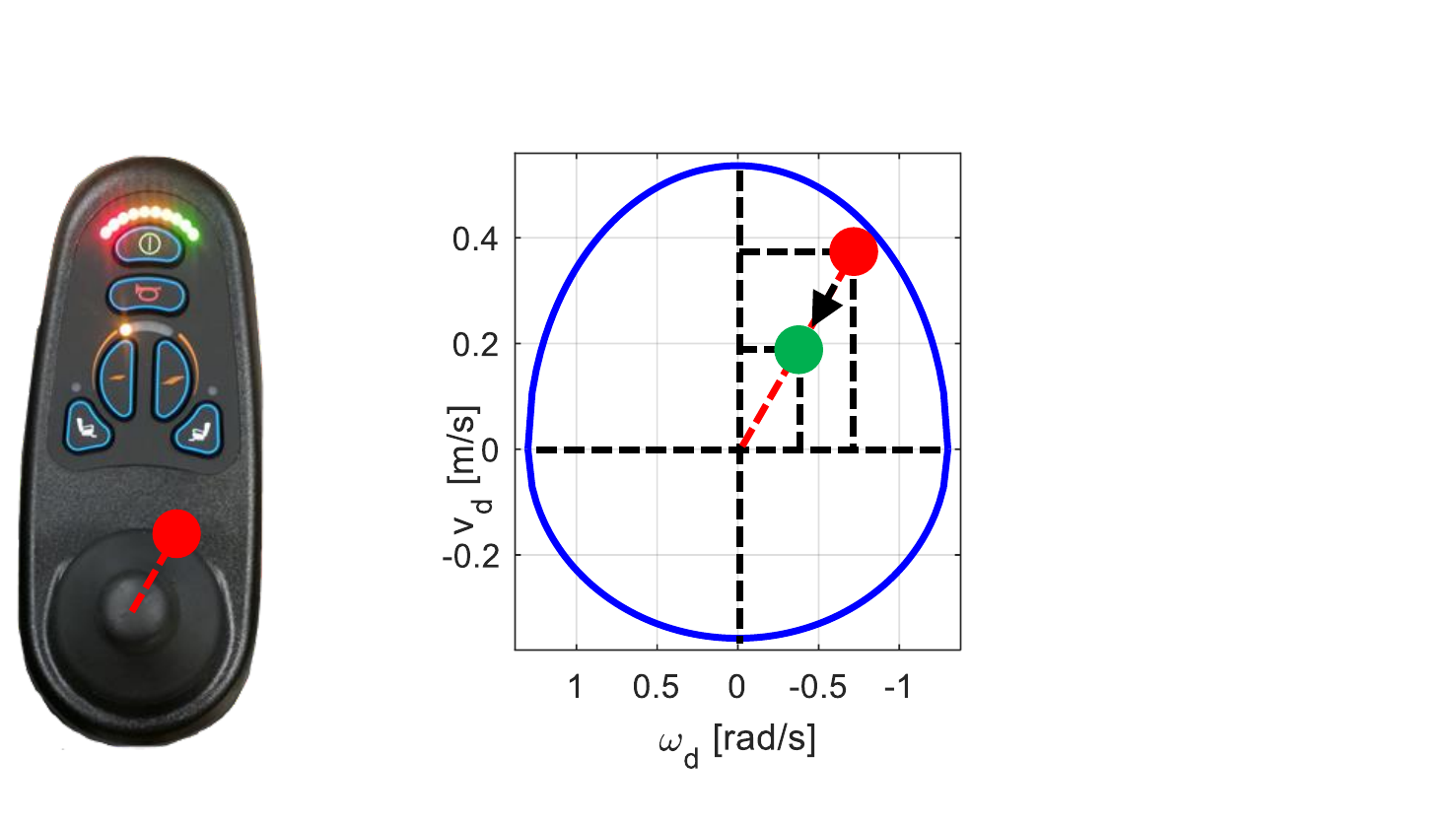}
   \caption{Applying ASC to joystick data: the unsafe user's demand (depicted by the \textcolor{red}{red} solid circle on the actual joystick - the left image, and on the joystick plane - the right image) is reduced linearly to a safe value (depicted by the \textcolor{green}{green} solid circle) before being sent as a request to the power module}
   \label{fig:whc_joyToVelo_rela}
\end{figure}

The power wheelchair is actuated via a standard wheelchair joystick. Its position (location) on the joystick plane is in a linear one-to-one relation to the output steady-state velocity of the vehicle. This is schematically illustrated in Fig.~\ref{fig:whc_joyToVelo_rela} where the position of the joystick (the \textcolor{red}{red} solid circle) can be mapped on the joystick velocities $(v_\text{d},\omega_\text{d})$-plane. This motivates the analysis on the $(v_\text{d},\omega_\text{d})$-plane hereafter. 
Say, the \textcolor{red}{red} solid circle in Fig.~\ref{fig:whc_joyToVelo_rela} corresponds to the actual user demand. If the current vehicle velocity $v$ falls within the ABC triangle of Fig.~\ref{fig:whc_schematic_ASC}, then the supervisory control (SC) needs to reduce it towards a safe value as explained in the previous section. By applying the same reduction ratio on both the $v_\text{d}$-axis as well as on the $\omega_\text{d}$-axis, will lead us to the \textcolor{green}{green} solid circle in Fig.~\ref{fig:whc_schematic_ASC}, which: (i) is safe from an obstacle avoidance point of view; (ii) represents $(v_\text{d}+v_\text{u}, \omega_\text{d}+\omega_\text{u})$ sent as a request to the power module (servo-drives) that actuate the vehicle's motors. Note that all this analysis is straightforward in steady-state conditions where $v_\text{d}\equiv{}v$ and $\omega_\text{d} \equiv \omega$; the transitory effects still need to tested in practice.   

An interesting fact of this semi-active (dampening) control design is that in all situations it will not contradict the user's intention (e.g. cases where the driver demands to advance to the left whereas the SC would make the vehicle turn right would not occur). This is due to the linear relation between the \textcolor{red}{red} solid circle and the \textcolor{green}{green} solid circle, both located on the \textcolor{red}{red} dashed line in Fig.~\ref{fig:whc_joyToVelo_rela}. Such conflicting situations were analysed theoretically in \cite{Tra2015} and this paper provides a practical solution to that. 

To end with, we provide a sample C++ ROS code of ASC that shall be used in the next section to carry out experiments. The interested reader can access it on 
\href{https://github.com/UCL-Aspire-Create/whc_pub_ASC_policy_iter.git}{https://\-git\-hub.com/UCL-As\-pi\-re-Cre\-ate/whc\_pub\_ASC\_po\-li\-cy\_i\-ter.git}

\section{Experiments}
\begin{figure}[!htb]
\centering
\vspace{1ex}
\subfloat[real world]{
 \mbox{  \includegraphics[scale=.19, trim=0ex 30ex 0ex 0ex, clip]{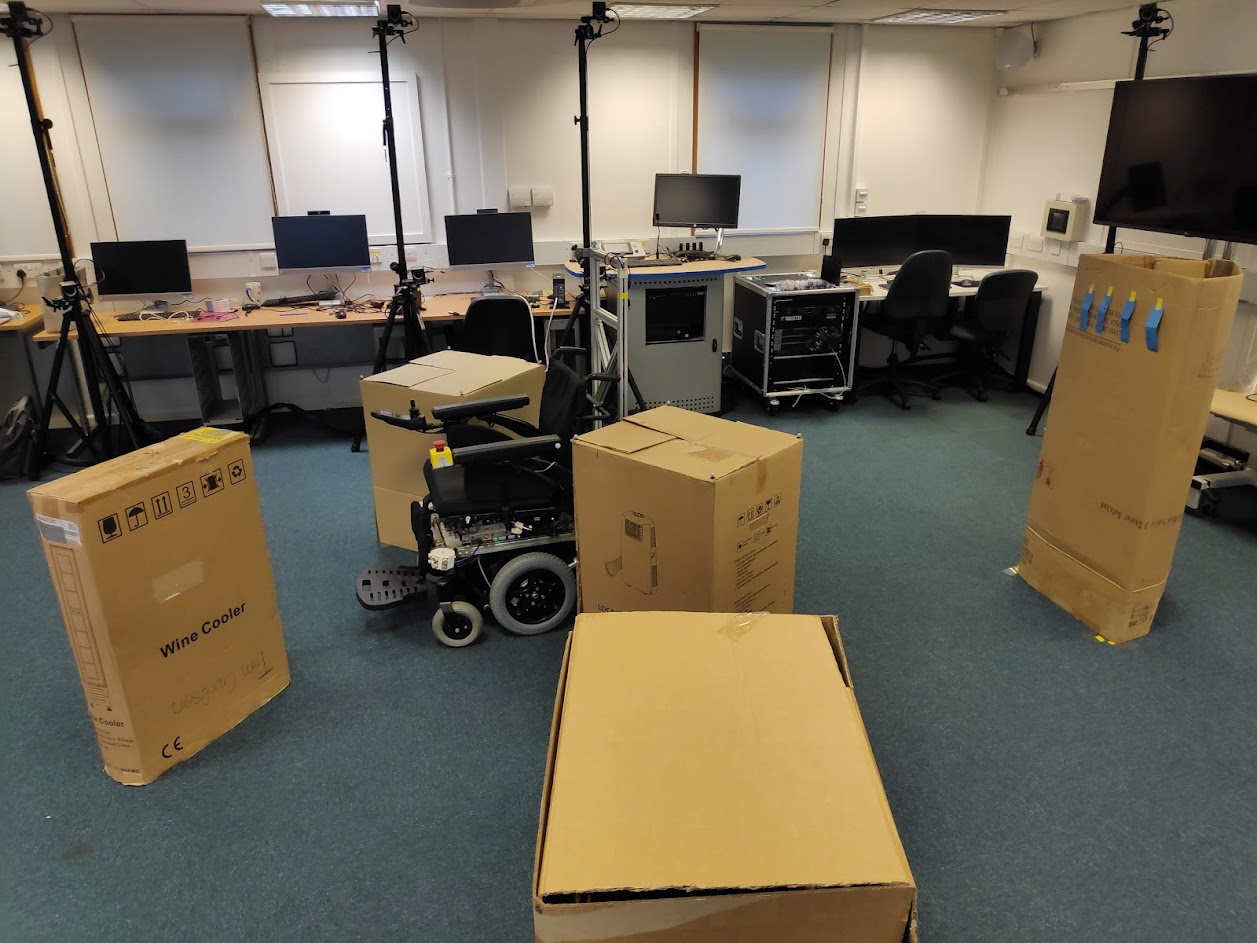} } 
 \label{fig:circuit_real}
} %
\hfill
\subfloat[real-time 3D simulation in Unity using Optitrack markers placed on obstacles and the wheelchair]{ 
 \mbox{ \includegraphics[scale=.3, trim=30ex 16ex 50ex 28ex, clip]{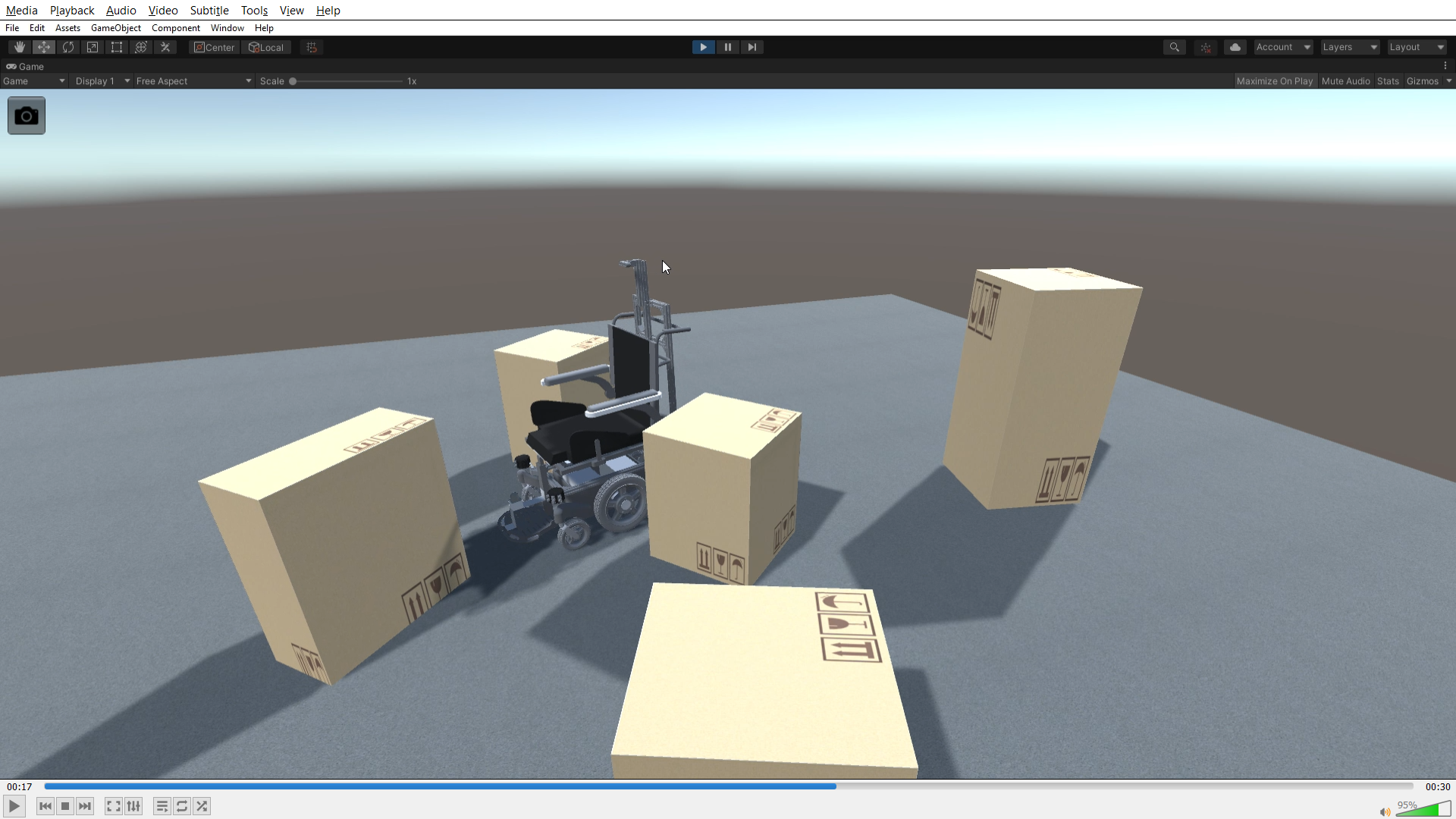} } 
 \label{fig:circuit_Unity} 
} %
\hfill
\subfloat[offline 2D simulation in Matlab (view from above); note the obstacles illustrated in \textcolor{orange}{orange}]{ 
 \mbox{ 
\includegraphics[scale=.73, trim=5ex .5ex 18.5ex 4.5ex, clip]{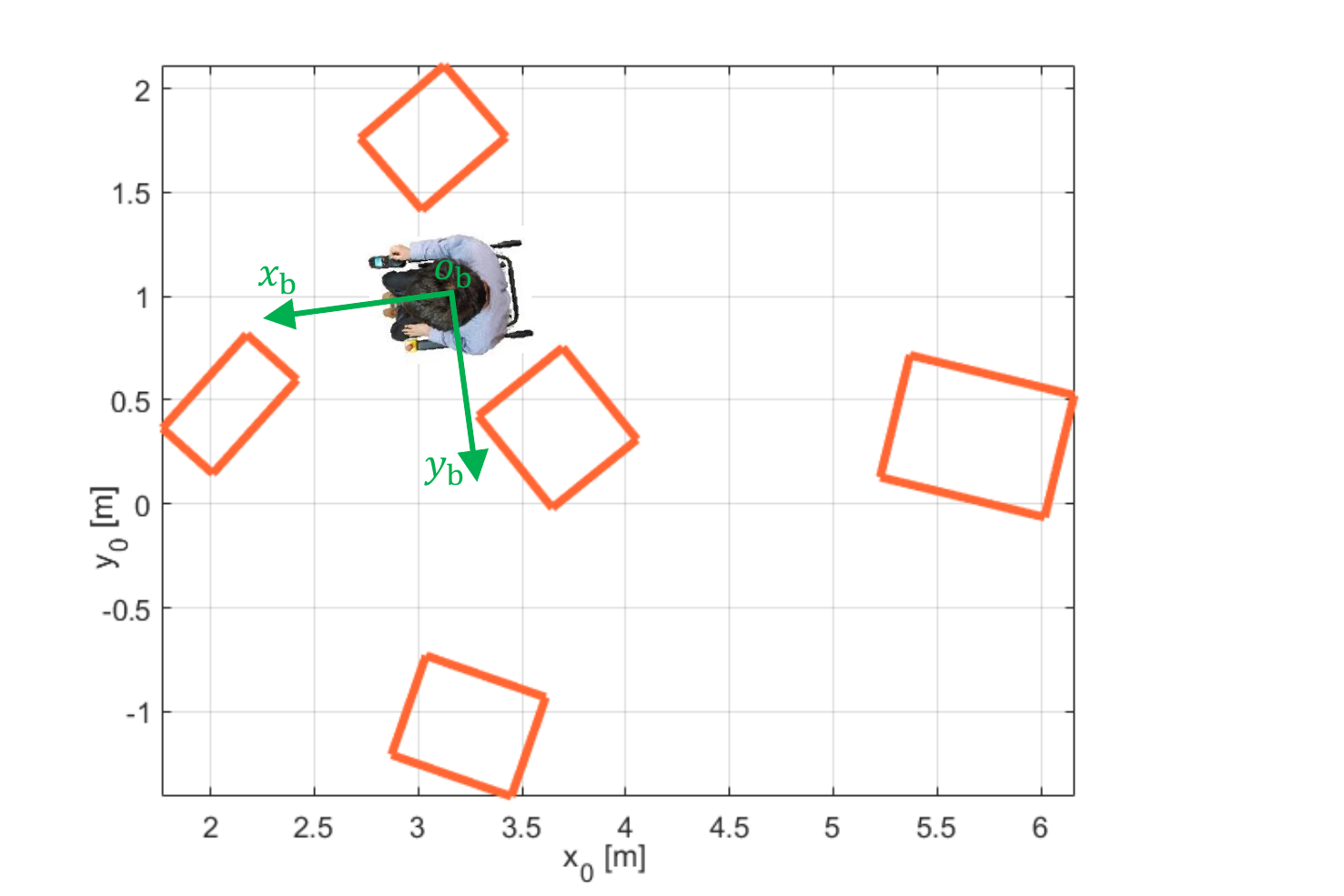}
} 
 \label{fig:circuit_Matlab} 
} %
\caption{The circuit used for testing the obstacle-avoidance algorithms. A video recording is available on \href{https://youtu.be/X-dQoLpYquQ}{https://you\-tu.be/X-dQoLpYquQ}}
\label{fig:circuit}
\end{figure}

To validate the proposed ASC, we created the circuit in Fig.~\ref{fig:circuit_real} with obstacles made of harmless cardboard boxes. 
We recruited four participants to this study, all able-bodied healthy people, fully trained researchers with experience in driving the wheelchair. From this point of view, their performance is closer to the \textit{expert driver} from the three stochastic driver models studied in this work.
Each participant was asked to execute the following operations: start driving the wheelchair from an initial position located to the left-hand side of Fig.~\ref{fig:circuit_real}, then drive between the obstacles, next pick one sticker attached to the cardboard box located to the right-hand side of Fig.~\ref{fig:circuit_real}, after that drive back between the obstacles to the starting point, next place the sticker in a basket and continue reiterating all these steps until all four blue stickers are placed inside the basket.
Drivers are allowed to advance forward only (not backwards).
This study is organized as a competition (contest) between these participants: the winner is the person who can drive in minimum time and with the least number of collisions. Say $t_\text{d}$ is the time to complete the track and $c_\text{d}$ is the number of collision, then each participant receives a score $s$ representing a \textit{penalized task completion time} measured in seconds, based on their performance:
\begin{align}
s = t_\text{d} + \alpha \, c_\text{d}
\end{align}
where $\alpha>0$ acts as a penalty for hitting obstacles. Inspired from circuit racing  \cite{FIA21}, we use $\alpha=5$ sec.

The aim of this study was to assess whether the ASC brings any benefit compared to: (i) the situation with no assistance (NA): this is used as the \textit{control group} - terminology according to the statistics community; (ii) a baseline assist-as-needed semi-autonomous control: here we shall use our previous rule-based (RB) algorithm \cite{TeZhCa2020b}. 
Hence, we address the \textit{research question}: how does the driving condition (NA, RB, ASC) influence the user performance?  
We make a first hypothesis that there is a difference between all these three driving conditions (NA, RB and ASC), and second that ASC is superior to RB. 
In order to reduce order effects (structural bias), the order of the driving conditions (NA, RB, ASC) is assigned randomly for each participant using \textit{simple random sampling} \cite[\S 3]{BrHo1977}.

An ethics approval was granted for this study by UCL Research Ethics Committee (ref. 6860/011).

\subsection{Data collection}
The interested reader can access a video recording on \href{https://youtu.be/X-dQoLpYquQ}{https://you\-tu.be/X-dQoLpYquQ}.
Table~\ref{tab:data-col} illustrates the data collected: each row represents a single participant's data; each column corresponds to one of the three driving conditions under consideration (NA, RB and ASC). 

\begin{table}[htbp]
\caption{Experimental data: time to complete the circuit in seconds and number of obstacles hit (the latter is indicated between round brackets) for 3 different driving conditions (NA = no assistance; RB = rule-based SC; ASC = Assist-as-needed SC).}
\label{tab:data-col}
\begin{center}
\begin{tabular}{|l|c|c|c|}
\hline
Participant &  NA     &  RB     & ASC       \\ \hline
1           & 111 (0) & 143 (0) & 132 (0)  \\ \hline 
2           & 125 (1) & 142 (0) & 136 (0)  \\ \hline 
3           & 102 (0) & 142 (0) & 133 (0)  \\ \hline 
4           & 110 (1) & 172 (1) & 148 (0)  \\ \hline 
\end{tabular}
\end{center}
\end{table}

\begin{figure}[htbp]
   \centering
   \includegraphics[scale=.62, trim=24ex 54ex 30ex 57ex, clip]{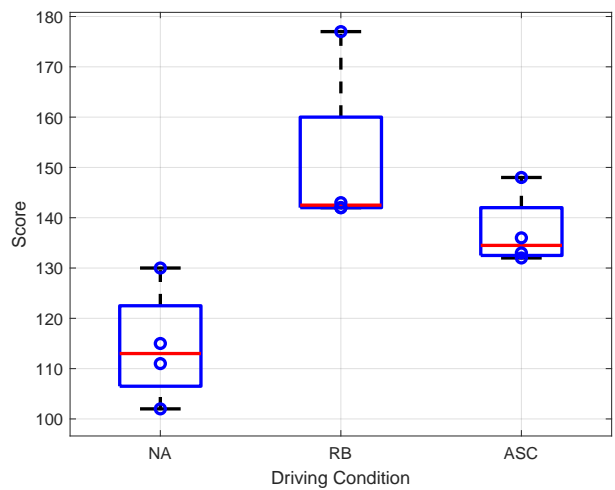}
   \caption{Box plots of the scores (lower is better)}
   \label{fig:boxplot}
\end{figure}

The experimental data in Table~\ref{tab:data-col} converted into scores used for this study can be visualized in Fig.~\ref{fig:boxplot}.  

\subsection{Statistical analysis}
In order to compare scores of the 3 aforementioned groups, we shall use ANOVA. Before deciding which ANOVA variant to use, first we checked the assumption of parametric data. Only the group NA meets this assumption (Shapiro-Wilk $p=0.836$, skewness $\rho=0.724$ and no outliers), whereas the other two groups violate this assumption: RB (Shapiro-Wilk $p=0.003$, skewness $\rho=2$ and no outliers); ASC (Shapiro-Wilk $p=0.125$, skewness $\rho=1.7$ and no outliers). This motivates the use of the One Way Repeated Measures Friedman's ANOVA which is a non-parametric alternative to the more popular parametric F-test.

Hence, as the data violated the parametric assumption, a Friedman's ANOVA -- a \textit{within participants} design \cite{NaFo2020}, was conducted to test whether the driving condition (NA, RB or ASC) affects the participants score in completing the track. The analysis revealed a significant main effect of driving condition ($\chi^2(2)=8$, $p=0.018$). The post-hoc analysis using pairwise comparisons (Durbin-Conover) revealed a significant difference ($p<.001$) between pairs of groups: (i) the NA (mean=115, $\sigma=11.7$) and RB (mean=151, $\sigma=17.3$); (ii) the NA and ASC (mean=137, $\sigma=7.37$); (iii) the RB and ASC. This suggests that ASC is superior to RB although both are outperformed by NC.


\subsection{Limitations of the study}
First, increasing or decreasing the penalty $\alpha$ on hitting obstacles, will have an immediate effect on the scores and thus the entire statistical analysis. For instance, inspired by and consistent with circuit racing practice \cite{FIA21}, another option would be $\alpha=10$ sec. Alternatively, more general and elaborate methods to adjust the weighting factor $\alpha$ are discussed in \cite{HoTeHe1998}.

Second, due to the ongoing coronavirus pandemic, we had only been able to recruit a rather limited number of participants to this study. Although results are statistically significant, these should be confirmed in a follow-up study involving more participants.

\subsection{Discussion}
The statistical analysis suggested that ASC outperforms RB, which is quite a positive result because it shows a progress in our research.
However, the scope of this result is limited to a particular circuit and does not account for other real-world scenarios (e.g. entering an elevator, advancing on a wide corridor with static obstacles, etc.) - see the standardized circuits in \cite{LeFrDe2020}. Follow-up studies are necessary to address these situations, including avoiding dynamic obstacles like pedestrians walking \cite{ZhAmEb2021}.

The rather intriguing fact that by disabling assistive control (NA group) the participants to this study obtained better scores compared to enabling assistive control (RB and ASC) can be explained by multiple factors. First, by the nature of these control solutions: both gradually take authority from the driver as the risk of collision increases. This translates into reduced effective velocities and thus the time to complete the circuit increases. Second, the participants to this study were experienced drivers (although not experts). Therefore they handled rather easily the obstacle avoidance task. Third, participants were all able-bodied whereas we expect assistive control to prove its real-world usefulness for people with known impairments (e.g. cognitive, visual, hearing, physical, etc.) \cite{LeFrDe2020}. To summarize, there is a trade-off between the risk of collision and the ability to finish fast the circuit: the price to pay for reducing the risk of collision when enabling assistive control is that it takes more time in average to complete the circuit.
In essence, from a broad perspective, our contribution has the potential to transform the lives of many people by creating a transportation system that empowers the user by means of technology to carry out safely everyday tasks (without bumping unintentionally into static obstacles).

Participants to this study appreciated the smoothness of ASC. However, the ASC might exhibit a more sudden (bumpy) behavior. This can be due to multiple factors: (i) the maps generated by the stochastic driver models used (recall Fig.~\ref{fig:whc_policy-iter-algo-cost-fcts}); (ii) this control design is based on the steady-state assumption that $v=v_\text{d}+v_\text{u}$ (see section~\ref{sec:ASC}), thus ignoring transitory effects. These aspects need to be investigated in future work.  

Our statistical analysis used a combined metric (a score), by weighting together the task completion time and the number of collisions. Instead, if we rerun the statistical analysis for each separate metric, the following results are obtained. First, the analysis using the task completion time revealed a similar (analogous) result: a significant effect of driving condition ($\chi^2(2)=8$, $p=0.018$). Second, the analysis using the number of collisions showed a non-significant effect of diving condition ($\chi^2(2)=3$, $p=0.223$), meaning we cannot conclude that ASC is superior to NA or RB: the fact that no collision occurred for ASC group in Table~\ref{tab:data-col}, contrary to the other groups, might be the result of a lucky sample; we need more participants to the study in order to obtain a statistically significant result in this setting.

\subsection{Software architecture}
All the sensors mounted on the research platform in Fig.~\ref{fig:instrum-whc} are responsible for building environment awareness capability for the robotic vehicle. For testing ASC we used: (i) the standard joystick to capture user's input $(v_\text{d},\omega_\text{d})$, (ii) the wheel encoders to estimate the actual velocity $v$, and (iii) the front-facing ultrasonic sensors to detect the nearby obstacles (see Fig.~\ref{fig:whc_under_seat}). 
Although not considered here, the alternative to use time-of-flight sensors placed under the footplate (see Fig.~\ref{fig:whc_pos_track} and \cite{MeTeWa2021}) is subject to future research.
The software architecture relies on a single-board computer (a Raspberry Pi 3b+) running Robot Operating System (ROS) on Ubuntu 16.04.

The simulators in Figs.~\ref{fig:circuit_Unity} and \ref{fig:circuit_Matlab} were used for visualization purpose.
Although not considered here, these simulators can be extended into virtual reality environments making use of \textit{digital twins} \cite{HaAu2021}, which is subject to future research.

\section{Conclusions}
In this article we developed and tested an assist-as-needed algorithm (we call it AssistMe) that provides a robotic vehicle with the intelligence (or capability) to avoid obstacles by collaborating with the user and thus ensuring a safe driving experience. We show how this algorithm can be personalized to meet the needs of each specific driver, based on their stochastic user model (we called them the expert driver, the blind driver and the naughty child); it requires low-cost hardware components, e.g. ultrasonic sensors, wheel encoders, a single-board computer, etc. The algorithm makes use of pre-computed optimal maps of the average time to hit an obstacle located ahead of the vehicle: we formulated a time-optimal stochastic shortest path problem and solved it numerically by implementing a computationally tractable algorithm relying on policy iteration.
An experimental study with healthy participants, showed an improved score of this assist-as-needed algorithm over a baseline rule-based control.

\section*{Acknowledgment}
We would like to thank George Walker for developing the Unity environment in Fig.~\ref{fig:circuit_Unity}; Marie Babel and her team (François Pasteau, Louise Devigne) at INSA and IRISA -- Rennes for providing us with an RNET controller used to: (i) publish standard joystick data on a ROS topic, and (ii) send \textit{virtual joystick} messages on the RNET bus to the power module.

\section*{Declaration of competing interest}
The authors declare that they have no known competing financial interests or personal relationships that could have appeared to influence the work reported in this paper. 

\section*{Funding}
This work is funded by the project INTERREG VA FMA ADAPT -- ``Assistive Devices for empowering disAbled People through robotic Technologies'' http://adapt-project.com/index.php. The FMA Program is a European Territorial Cooperation program which aims to finance ambitious cooperation projects in the border region between France and England. The Program is supported by the European Regional Development Fund (FEDER).


\end{document}